\documentclass[11pt,a4paper]{article}

\usepackage{jheppub}

\usepackage{amsmath}
\allowdisplaybreaks
\usepackage{autobreak}
\usepackage{slashed}

\usepackage{array}
\newcolumntype{T}{>{\ttfamily} c}
\newcolumntype{M}{>{$\displaystyle} c <{$}}

\usepackage{axodraw2}

\newcommand{\MS}{{\ensuremath{\text{MS}}}}
\newcommand{\MSbar}{{\ensuremath{\overline{\text{MS}}}}}
\newcommand{\MOM}{MOM}
\newcommand{\MOMtilde}{{\ensuremath{\widetilde{\text{MOM}}}}}
\newcommand{\MiniMOM}{MiniMOM}
\newcommand{\MM}{\text{MM}}

\newcommand{\Forcer}{{\sc Forcer}}
\newcommand{\Mincer}{{\sc Mincer}}
\newcommand{\FORM}{{\sc Form}}

\newcommand{\ep}{\varepsilon}
\newcommand{\als}{\alpha_{\rm s}}

\newcommand{\colorfactor}[1]{\textcolor{Blue}{#1}}
\newcommand{\gaugefactor}[1]{\textcolor{Red}{#1}}
\newcommand{\exprname}{}
\newcommand{\exprfile}{}
\newcommand{\exprlabel}{} \let\exprlabel=\empty

{
  \catcode`\^^M=\active
  
}
\newcommand{\putresultb}{}
{
  \catcode`\^^M=\active
  \gdef\putresultb{
    \begin{align}
      \begin{autobreak}
        \MoveEqLeft
        \exprname{0} =
        \frac{11}{3} \colorfactor{\ca}
-\frac{4}{3} \colorfactor{\tf \nf}
 ,
      \end{autobreak} \nonumber \\[2mm]
      \begin{autobreak}
        \MoveEqLeft
        \exprname{1} =
         ,
      \end{autobreak} \nonumber \\[2mm]
      \begin{autobreak}
        \MoveEqLeft
        \exprname{2} =
         ,
      \end{autobreak} \nonumber \\[2mm]
      \begin{autobreak}
        \MoveEqLeft
        \exprname{3} =
         ,
      \end{autobreak} \nonumber \\[2mm]
      \begin{autobreak}
        \MoveEqLeft
        \exprname{4} =
         .
      \end{autobreak}
      \ifx\exprlabel\empty
      \else
        \label{res:\exprlabel{}}
      \fi
    \end{align}
  }
}

\makeatletter
\newcommand{\xil}{}
\newcommand{\xilpow}{}
\def\xil{\@ifnextchar^{\xilpow}{\xilpow^\empty}}%
\def\xilpow^#1{%
  \ifx#1\empty
    {\xi}%
  \else
    {\xi^{\,#1}}%
  \fi
}
\newcommand{\z}{}
\newcommand{\znpow}{}
\def\z#1{\@ifnextchar^{\znpow{#1}}{\znpow{#1}^\empty}}%
\def\znpow#1^#2{%
  \ifx#2\empty
    {\zeta_{#1}}%
  \else
    {\zeta_{#1}^{\,#2}}%
  \fi
}
\newcommand{\ca}{}
\newcommand{\capow}{}
\def\ca{\@ifnextchar^{\capow}{\capow^\empty}}%
\def\capow^#1{%
  \ifx#1\empty
    {C_{\!A}^{}}%
  \else
    {C_{\!A}^{\,#1}}%
  \fi
}
\newcommand{\cf}{}
\newcommand{\cfpow}{}
\def\cf{\@ifnextchar^{\cfpow}{\cfpow^\empty}}%
\def\cfpow^#1{%
  \ifx#1\empty
    {C_{F}^{}}%
  \else
    {C_{F}^{\,#1}}%
  \fi
}
\newcommand{\tf}{}
\newcommand{\tfpow}{}
\def\tf{\@ifnextchar^{\tfpow}{\tfpow^\empty}}%
\def\tfpow^#1{%
  \ifx#1\empty
    {T_{\!F}^{}}%
  \else
    {T_{\!F}^{\,#1}}%
  \fi
}
\newcommand{\nf}{}
\newcommand{\nfpow}{}
\def\nf{\@ifnextchar^{\nfpow}{\nfpow^\empty}}%
\def\nfpow^#1{%
  \ifx#1\empty
    {n_{\!f}^{}}%
  \else
    {n_{\!f}^{\,#1}}%
  \fi
}
\def\dfAAna{{\frac{d_A^{\,abcd}d_A^{\,abcd}}{N_{\!A}}}}
\def\dfFAna{{\frac{d_F^{\,abcd}d_A^{\,abcd}}{N_{\!A}}}}
\def\dfFFna{{\frac{d_F^{\,abcd}d_F^{\,abcd}}{N_{\!A}}}}

\makeatother

\makeatletter
\def\@fpheader{\relax}
\makeatother

\preprint{Nikhef 2017-013 \\ \mbox{} \hfill LTH 1125}

\title{Four-loop QCD propagators and vertices with one vanishing external 
momentum}

\author[a,b]{B.~Ruijl,}
\author[a]{T.~Ueda,}
\author[a]{J.A.M.~Vermaseren}
\author[c]{and A.~Vogt}

\affiliation[a]{%
  Nikhef Theory Group, \\
  Science Park 105, 1098 XG Amsterdam,
  The Netherlands}
\affiliation[b]{%
  Leiden Centre of Data Science, Leiden University, \\
  Niels Bohrweg 1, 2333 CA Leiden,
  The Netherlands}
\affiliation[c]{%
  Department of Mathematical Sciences, University of Liverpool, \\
  Liverpool L69 3BX, United Kingdom}

\emailAdd{benrl@nikhef.nl}
\emailAdd{tueda@nikhef.nl}
\emailAdd{t68@nikhef.nl}
\emailAdd{andreas.vogt@liverpool.ac.uk}

\abstract{%
  We have computed the self-energies and a set of three-particle vertex 
  functions for massless QCD at the four-loop level in the \MSbar{} 
  renormalization scheme.
  The vertex functions are evaluated at points where one of the momenta
  vanishes.
  Analytical results are obtained for a generic gauge group and with the 
  full gauge dependence, which was made possible by extensive use of the
  \Forcer{} program for massless four-loop propagator integrals.
  The bare results in dimensional regularization are provided in terms of 
  master integrals and rational coefficients; the latter are exact in any 
  space-time dimension.
  Our results can be used for further precision investigations of the 
  perturbative behaviour of the theory in schemes other than \MSbar{}.
  As an example, we derive the five-loop beta function in a relatively 
  common alternative, the minimal momentum subtraction (\MiniMOM{}) scheme.
}


\begin{document}
\maketitle
\flushbottom

\setlength{\parskip}{0.1cm plus 0.3cm}

\vspace*{-1mm}
\section{Introduction}

Motivated by the goal to obtain predictions that are as precise as possible
for quantities entering benchmark high-energy processes, there have been 
significant recent developments of computational techniques and methodologies 
in higher-order perturbative calculations for Quantum Chromodynamics (QCD)\@.
As a result,
computations of the basic renormalization group functions of QCD~%
\cite{%
Gross:1973id,%
Politzer:1973fx,%
Caswell:1974gg,%
Jones:1974mm,%
Egorian:1978zx,%
Tarrach:1980up,%
Tarasov:1980au,%
Tarasov:1982gk,%
Larin:1993tp,%
vanRitbergen:1997va,%
Chetyrkin:1997dh,%
Vermaseren:1997fq,%
Chetyrkin:2004mf,%
Czakon:2004bu%
}
have recently reached the five-loop level~%
\cite{%
Baikov:2014qja,%
Baikov:2016tgj,%
Luthe:2016ima,%
Luthe:2016xec,%
Herzog:2017ohr,%
Luthe:2017ttc,%
Baikov:2017ujl%
}.
In the modified minimal subtraction (\MSbar{}) scheme~%
\cite{tHooft:1973mfk,Bardeen:1978yd},
the standard renormalization scheme for perturbative QCD which is closely 
related to dimensional regularization~%
\cite{Bollini:1972ui,tHooft:1972tcz},
these functions are obtained by computing single poles in corresponding
Green's functions.
Since only the poles are required, the above four- and five-loop results were 
first obtained using the method of infrared rearrangement~%
\cite{%
Vladimirov:1979zm,%
Chetyrkin:1982nn,%
Chetyrkin:1984xa,%
Chetyrkin:2017ppe,%
Herzog:2017bjx,%
Chetyrkin:1996sr,%
Chetyrkin:1996ez,%
Misiak:1994zw,%
Chetyrkin:1997fm%
}
which simplifies computations without changing the ultraviolet singular
structure, but modifies the finite parts.

\newpage

The main aim of this work is to provide the self-energies and a set of 
vertices with one vanishing external momentum for massless QCD at the 
four-loop level.
The unrenormalized results are exact in terms of $\ep =(D-4)/2$, where 
$D$ is the space-time dimension, and four-loop master integrals~%
\cite{Baikov:2010hf,Lee:2011jt}.
The renormalized $D\!=\!4$ results are given in the \MSbar{} scheme.
The computation has been performed for a general gauge group and in
an arbitrary covariant linear gauge, by using the \Forcer{} program~%
\cite{Ueda:2016sxw,Ueda:2016yjm,FORCER}
for massless four-loop propagator-type integrals.
For the vertices, setting one of the momenta to zero effectively reduces 
vertex integrals to propagator-type integrals. In QCD this does not create 
IR divergences, which means the poles do not change.
At the three-loop level, similar computations were performed in ref.~%
\cite{Chetyrkin:2000dq}, but with an expansion in $\ep$.
In addition, studies of QCD vertices in perturbation theory for various
configurations include refs.~%
\cite{%
Celmaster:1979km,%
Ball:1980ax,%
Pascual:1980yu,%
Braaten:1981dv,%
Davydychev:1996pb,%
Davydychev:2001uj,%
Binger:2006sj,%
Davydychev:1997vh,%
Davydychev:1998aw,%
Davydychev:2000rt,%
Chetyrkin:2000fd,%
Gracey:2011vw,%
Gracey:2014mpa%
}.

We compute all QCD vertices in a general linear covariant gauge, with the 
exception of the four-gluon vertex for which there are at least three 
difficulties:
first, two momenta have to be nullified before the diagrams become 
propagator-like.
Second, the number of diagrams is large at four loops.
Third, the colour structure for a generic group is no longer an overall factor, 
but will be term dependent. Additionally, the renormalization constant
is completely determined through the Slavnov-Taylor identities~%
\cite{Taylor:1971ff,Slavnov:1972fg}, which means the quartic gluon
vertex does not provide extra information in this context. 

A direct application of our results is to compute conversion factors from
the \MSbar{} scheme to momentum subtraction schemes, see, e.g.,~refs.~%
\cite{Celmaster:1979km,Braaten:1981dv}, for renormalization group functions.
Unlike the \MSbar{} scheme, momentum subtraction schemes are defined in
a regularization-independent way.
In these schemes, the field renormalizations are performed such that finite 
radiative corrections on propagators are absorbed as well as divergences and 
hence they coincide with their tree-level values at the renormalization point.
Then one of (or an arbitrary linear combination of) the vertex functions
is normalized to its tree-level value and the other vertices are fixed via the 
Slavnov-Taylor identities.
Common choices for the subtraction point of the vertex are a symmetric point
(referred as \MOM{} schemes) and an asymmetric point where one of the momenta 
is nullified, sometimes referred as \MOMtilde{} schemes.
The latter choice corresponds to our result for the vertex functions.
Indeed, ref.~\cite{Chetyrkin:2000dq} derived four-loop beta functions in
four particular \MOMtilde{} schemes from that in the \MSbar{} scheme
by computing conversion factors via finite parts of two- and three-point
functions in the \MSbar{} scheme.

As an example application, we provide the five-loop beta function in the 
minimal momentum subtraction (\MiniMOM{}) scheme introduced in ref.~%
\cite{vonSmekal:2009ae}, thus extending previous results~%
\cite{vonSmekal:2009ae,Gracey:2013sca} by one order in the coupling constant. 
This scheme, see the preceeding references for a detailed discussion, is 
more convenient than \MSbar{} for extending analyses of the strong coupling 
constant and its scale dependence into the non-perturbative regime, e.g., 
via lattice QCD; for a recent analysis see ref.~\cite{Ayala:2017tco}. 
In the perturbative regime the \MiniMOM{} scheme provides an alternative to 
\MSbar{} for studying the behaviour and truncation uncertainty of the 
perturbation series for benchmark quantities such as the $R$-ratio in 
$e^+e^-$ annihilation and the Higgs-boson decay to gluons, see refs.~%
\cite{Gracey:2014pba,Kataev:2015yha,Zeng:2015gha}.
%


The remainder of this article is organized as follows:
In section~\ref{sect:prelim} we specify our notations for the self-energies 
and vertex functions and recall their renormalization.
In~section~\ref{sect:compt} we address technical details and checks of the
computation.
Due to their length, the results for the renormalized self-energies and vertex
functions for a generic gauge group and a general linear covariant gauge are
deferred to appendix \ref{sec:results}\@. 
In section~\ref{sect:betaMM} the five-loop beta function is presented in the 
\MiniMOM{} scheme for QCD in the Landau gauge; the general result and the 
corresponding \MSbar{}-to-\MiniMOM{} conversion factor for the coupling
constant are provided in appendix \ref{sec:mmresults}\@.
In section~\ref{sect:conclusion} we summarize and give a brief outlook.

Our result can be obtained in a digital form as ancillary files of this article
on the preprint server \url{https://arXiv.org}.
They are also available from the authors upon request.
The files contain the bare results for the self-energies and vertices in terms 
of master integrals with coefficients that are exact for any dimension $D$, 
as well as the results in the \MSbar{} scheme for $D=4$. The notations in these
files can be found in appendix~\ref{sec:notationfile}\@.

\section{Preliminaries}
\label{sect:prelim}

We first summarize the notations for self-energies and vertex
functions with one vanishing momentum presented in this article.
In most cases we follow the conventions in ref.~\cite{Chetyrkin:2000dq}.%
\footnote{%
  We note that these conventions may be different from the ones commonly used
  in the literature. In fact, the Feynman rules in
  \Forcer{} are different as well, and hence we occasionally had to convert
  intermediate results from one convention to the other and back.
}

\subsection{Self energies}

\begin{figure}
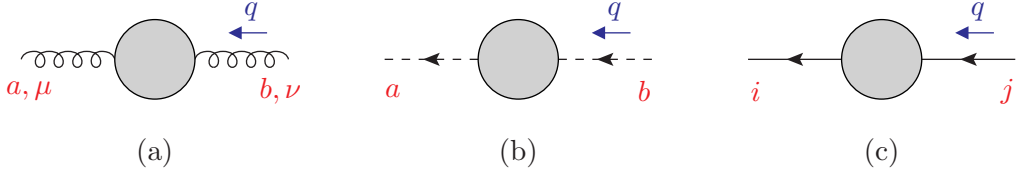

  \centering
  \setlength{\tabcolsep}{15pt}
  \begin{tabular}{ccc}
    \begin{axopicture}(106,46)(-53,-23)%
      \Gluon(-50,0)(-15,0){3}{4}%
      \Gluon(15,0)(50,0){3}{4}%
      \GCirc(0,0){15}{0.82}%
      \SetArrowInset{0}%
      \SetArrowAspect{1}%
      \SetColor{Blue}%
      \LongArrow(42,10)(30,10)%
      \Text(36,18){$q$}%
      \SetColor{Red}%
      \Text(-47,-12){$a,\mu$}%
      \Text(47,-12){$b,\nu$}%
    \end{axopicture}
    &
    \begin{axopicture}(106,46)(-53,-23)%
      \DashArrowLine(50,0)(15,0){3}%
      \DashArrowLine(-15,0)(-50,0){3}%
      \GCirc(0,0){15}{0.82}%
      \SetArrowInset{0}%
      \SetArrowAspect{1}%
      \SetColor{Blue}%
      \LongArrow(42,10)(30,10)%
      \Text(36,18){$q$}%
      \SetColor{Red}%
      \Text(-47,-12){$a$}%
      \Text(47,-12){$b$}%
    \end{axopicture}
    &
    \begin{axopicture}(106,46)(-53,-23)%
      \ArrowLine(50,0)(15,0)%
      \ArrowLine(-15,0)(-50,0)%
      \GCirc(0,0){15}{0.82}%
      \SetArrowInset{0}%
      \SetArrowAspect{1}%
      \SetColor{Blue}%
      \LongArrow(42,10)(30,10)%
      \Text(36,18){$q$}%
      \SetColor{Red}%
      \Text(-47,-12){$i$}%
      \Text(47,-12){$j$}%
    \end{axopicture}
    \\ (a) & (b) & (c)
  \end{tabular}
  \caption{%
    The gluon, ghost and quark self-energies $\Pi^{ab}_{\mu\nu}(q)$ (a),
    $\tilde{\Pi}^{ab}(q)$ (b) and $\Sigma^{ij}(q)$ (c).
  }
  \label{fig:self-energy}
\end{figure}

The gluon, ghost and quark self-energies (figure~\ref{fig:self-energy})
are of the form
\begin{align}
\label{firstdef}
  \Pi^{ab}_{\mu\nu}(q)
    &\:=\: - \delta^{ab} (q^2 g_{\mu\nu} - q_\mu q_\nu) \Pi(q^2) , \\[1mm]
  \tilde{\Pi}^{ab}(q)
  &\:=\: \delta^{ab} q^2 \tilde{\Pi}(q^2) , \\[1mm]
  \Sigma^{ij}(q)
  &\:=\: \delta^{ij} \slashed{q} \Sigma_V(q^2) .
\end{align}
The colour indices are understood such that $a$ and $b$ are for the adjoint
representation of the gauge group, $i$ and $j$ for the representation to which
the quark belongs.
The form factors $\Pi(q^2)$, $\tilde{\Pi}(q^2)$ and $\Sigma_V(q^2)$ can
easily be extracted from contributions of the corresponding one-particle
irreducible diagrams by applying projection operators~\cite{Chetyrkin:2000dq}
(the same holds for the vertex functions discussed below).
They are related to the full gluon, ghost and quark propagators as follows:
\begin{align}
  D^{ab}_{\mu\nu}(q)
  &\:=\: \frac{\delta^{ab}}{-q^2} \biggl[
     \Bigl( - g_{\mu\nu} + \frac{q_\mu q_\nu}{q^2} \Bigr)\,\frac{1}{1+\Pi(q^2)}
     - \xi \,\frac{q_\mu q_\nu}{q^2}
     \biggr] , \\
  \Delta^{ab}(q)
  &\:=\: \frac{\delta^{ab}}{-q^2} \,\frac{1}{1+\tilde{\Pi}(q^2)} \:, \\
  S^{ij}(q)
  &\:=\: \frac{\delta^{ij}}{-q^2} \,\frac{\slashed{q}}{1+\Sigma_V(q^2)} \:.
\end{align}
Here the Landau gauge corresponds to $\xi=0$, and the Feynman gauge to $\xi=1$.
 We note that this convention differs from that in the widely used \FORM{} 
version \cite{Larin:1991fz} of the \Mincer{} program~\cite{Gorishnii:1989gt} 
for three-loop self-energies, where the symbol {\tt xi} represents $1 - \xi$.

\subsection{Triple-gluon vertex}

\begin{figure}[b]
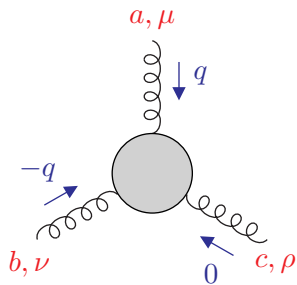

  \vspace{0.4cm}
  \centering
  \begin{axopicture}(106,100)(-53,-40)%
    \Gluon(0,50)(0,15){3}{4}%
    \Gluon(-43,-25)(-13,-8){3}{4}%
    \Gluon(43,-25)(13,-8){3}{4}%
    \GCirc(0,0){15}{0.82}%
    \SetArrowInset{0}%
    \SetArrowAspect{1}%
    \SetColor{Blue}%
    \LongArrow(10,42)(10,30)%
    \LongArrow(-40.4,-12)(-30,-6)%
    \LongArrow(30.4,-31)(20,-25)%
    \Text(18,37){$q$}%
    \Text(-43,1){$-q$}%
    \Text(22,-38){$0$}%
    \SetColor{Red}%
    \Text(0,58){$a,\mu$}%
    \Text(-46,-34){$b,\nu$}%
    \Text(46,-34){$c,\rho$}%
  \end{axopicture}
  \caption{The triple-gluon vertex with one vanishing momentum,
           $\Gamma^{abc}_{\mu\nu\rho}(q,-q,0)$.}
  \label{fig:ggg}
\end{figure}

Without loss of generality, one can set the momentum of the third gluon to 
zero, as depicted in figure~\ref{fig:ggg}.
Then the triple-gluon vertex can be written in the following form:
\begin{equation}
  \Gamma^{abc}_{\mu\nu\rho}(q,-q,0)
  \,=\,
  - i g f^{abc} \biggl[
    (2 g_{\mu\nu} q_\rho - g_{\mu\rho} q_\nu - g_{\rho\nu} q_\mu)\, T_1(q^2)
    - \Bigl( g_{\mu\nu} - \frac{q_\mu q_\nu}{q^2} \Bigr)\, q_\rho T_2(q^2)
  \biggr ] ,
  \label{eq:3g}
\end{equation}
where $g$ is the coupling constant and $f^{abc}$ are the structure constants of
the gauge group.
The first term in the square bracket corresponds to the tree-level vertex
while the second term arises from radiative corrections,
i.e., at the tree-level the form factors $T_{1,2}(q^2)$ read
\begin{equation}
  T_1(q^2) \bigl|_\text{tree} \,=\, 1 , \qquad
  T_2(q^2) \bigl|_\text{tree} \,=\, 0 .
\end{equation}

Because of Furry's theorem~\cite{Furry:1937zz} and the fact that we have no 
  colour-neutral particles, symmetric invariants with an odd number of 
  indices cannot occur for internal fermion lines. Neither can such 
  invariants occur for the adjoint representation. Hence, if we project out a 
  $d^{abc}$ structure, we would get a scalar invariant with an odd number of 
  $f$ tensors, and such a combination must be zero. This has been checked 
  explicitly to the equivalent of six-loop vertices in 
  ref~\cite{vanRitbergen:1998pn}.
Due to the bosonic property of gluons, the totally antisymmetric colour factor
$f^{abc}$ leads to antisymmetric Lorentz structure as in eq.~\eqref{eq:3g}.
  One could consider another Lorentz structure,
  \begin{equation}
    - i g f^{abc} q_\mu q_\nu q_\rho T_3(q^2) .
    \label{eq:T3}
  \end{equation}
  However, a Slavnov-Taylor identity requires $T_3(q^2)$ to vanish.

\subsection{Ghost-gluon vertex}

\begin{figure}[t]
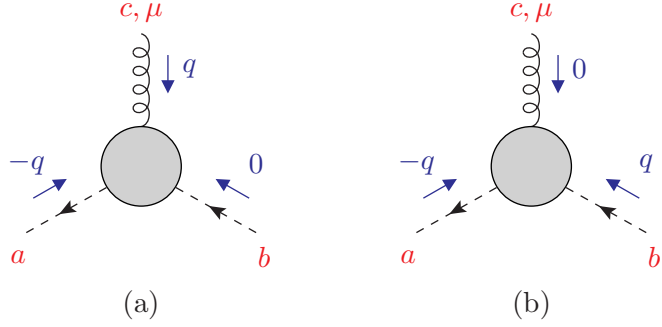

  \centering
  \setlength{\tabcolsep}{20pt}
  \begin{tabular}{cc}
    \begin{axopicture}(106,100)(-53,-40)%
      \Gluon(0,50)(0,15){3}{4}%
      \DashArrowLine(-13,-8)(-43,-25){3}%
      \DashArrowLine(43,-25)(13,-8){3}%
      \GCirc(0,0){15}{0.82}%
      \SetArrowInset{0}%
      \SetArrowAspect{1}%
      \SetColor{Blue}%
      \LongArrow(10,42)(10,30)%
      \LongArrow(-40.4,-12)(-30,-6)%
      \LongArrow(40.4,-12)(30,-6)%
      \Text(18,37){$q$}%
      \Text(-43,1){$-q$}%
      \Text(43,1){$0$}%
      \SetColor{Red}%
      \Text(0,58){$c,\mu$}%
      \Text(-46,-34){$a$}%
      \Text(46,-34){$b$}%
    \end{axopicture}
    &
    \begin{axopicture}(106,100)(-53,-40)%
      \Gluon(0,50)(0,15){3}{4}%
      \DashArrowLine(-13,-8)(-43,-25){3}%
      \DashArrowLine(43,-25)(13,-8){3}%
      \GCirc(0,0){15}{0.82}%
      \SetArrowInset{0}%
      \SetArrowAspect{1}%
      \SetColor{Blue}%
      \LongArrow(10,42)(10,30)%
      \LongArrow(-40.4,-12)(-30,-6)%
      \LongArrow(40.4,-12)(30,-6)%
      \Text(18,37){$0$}%
      \Text(-43,1){$-q$}%
      \Text(43,1){$q$}%
      \SetColor{Red}%
      \Text(0,58){$c,\mu$}%
      \Text(-46,-34){$a$}%
      \Text(46,-34){$b$}%
    \end{axopicture}
    \\ (a) & (b)
  \end{tabular}
  \caption{%
    The ghost-gluon vertex:
    (a) $\tilde{\Gamma}^{abc}_\mu(-q,0;q)$ with the vanishing incoming ghost
    momentum and (b) $\tilde{\Gamma}^{abc}_\mu(-q,q;0)$ with the vanishing gluon
    momentum.
  }
  \label{fig:hhg}
\end{figure}

Since the tree-level vertex is proportional to the outgoing ghost momentum,
nullifying this momentum gives identically zero in perturbation theory.
Therefore, we only have two possibilities to set one of the external momenta to zero.
One is the incoming ghost momentum and the other is the gluon momentum
(figure~\ref{fig:hhg}):
\begin{align}
  \tilde{\Gamma}_\mu^{abc}(-q,0;q)
    &\:=\: - i g f^{abc} q_\mu \tilde{\Gamma}_h(q^2) , \\[1mm]
  \tilde{\Gamma}_\mu^{abc}(-q,q;0)
    &\:=\: - i g f^{abc} q_\mu \tilde{\Gamma}_g(q^2) .
\end{align}
The subscript `$h$' of $\tilde{\Gamma}_h(q^2)$ indicates the function with
vanishing incoming ghost momentum, whereas `$g$' of $\tilde{\Gamma}_g(q^2)$
denotes the vanishing gluon momentum.
These functions are equal to one at the tree-level,
\begin{equation}
  \tilde{\Gamma}_h(q^2) \bigl|_\text{tree} \,=\,
  \tilde{\Gamma}_g(q^2) \bigl|_\text{tree} \,=\, 1.
\end{equation}

\subsection{Quark-gluon vertex}

\begin{figure}[b]
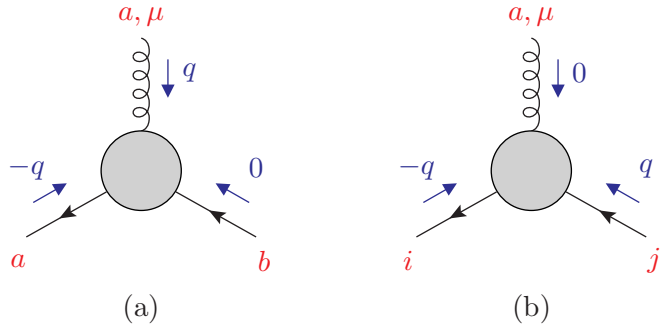

  \vspace{0.4cm}
  \centering
  \setlength{\tabcolsep}{20pt}
  \begin{tabular}{cc}
    \begin{axopicture}(106,100)(-53,-40)%
      \Gluon(0,50)(0,15){3}{4}%
      \ArrowLine(-13,-8)(-43,-25)%
      \ArrowLine(43,-25)(13,-8)%
      \GCirc(0,0){15}{0.82}%
      \SetArrowInset{0}%
      \SetArrowAspect{1}%
      \SetColor{Blue}%
      \LongArrow(10,42)(10,30)%
      \LongArrow(-40.4,-12)(-30,-6)%
      \LongArrow(40.4,-12)(30,-6)%
      \Text(18,37){$q$}%
      \Text(-43,1){$-q$}%
      \Text(43,1){$0$}%
      \SetColor{Red}%
      \Text(0,58){$a,\mu$}%
      \Text(-46,-34){$a$}%
      \Text(46,-34){$b$}%
    \end{axopicture}
    &
    \begin{axopicture}(106,100)(-53,-40)%
      \Gluon(0,50)(0,15){3}{4}%
      \ArrowLine(-13,-8)(-43,-25)%
      \ArrowLine(43,-25)(13,-8)%
      \GCirc(0,0){15}{0.82}%
      \SetArrowInset{0}%
      \SetArrowAspect{1}%
      \SetColor{Blue}%
      \LongArrow(10,42)(10,30)%
      \LongArrow(-40.4,-12)(-30,-6)%
      \LongArrow(40.4,-12)(30,-6)%
      \Text(18,37){$0$}%
      \Text(-43,1){$-q$}%
      \Text(43,1){$q$}%
      \SetColor{Red}%
      \Text(0,58){$a,\mu$}%
      \Text(-46,-34){$i$}%
      \Text(46,-34){$j$}%
    \end{axopicture}
    \\ (a) & (b)
  \end{tabular}
  \caption{%
    The quark-gluon vertex:
    (a) $\Lambda^a_{\mu,ij}(-q,0;q)$ with the vanishing incoming quark momentum
    and (b) $\Lambda^a_{\mu,ij}(-q,q;0)$ with the vanishing gluon momentum.
  }
  \label{fig:qqg}
\end{figure}

We consider the case of a vanishing incoming quark momentum and
the case of a vanishing gluon momentum (figure~\ref{fig:qqg}).
Nullifying the outgoing quark momentum gives the same result as nullifying
the incoming quark momentum.
Then the vertex can be written as
\begin{align}
  \Lambda^a_{\mu,ij}(-q,0;q)
    &\:=\: g T^a_{ij} \Bigl[
      \gamma_\mu \Lambda_q(q^2)
      + \gamma^\nu \biggl( g_{\mu\nu} - \frac{q_\mu q_\nu}{q^2} \biggr)
        \Lambda_q^T(q^2)
    \Bigr] , \\
  \Lambda^a_{\mu,ij}(-q,q;0)
    &\:=\: g T^a_{ij} \Bigl[
      \gamma_\mu \Lambda_g(q^2)
      + \gamma^\nu \biggl( g_{\mu\nu} - \frac{q_\mu q_\nu}{q^2} \biggr)
        \Lambda_g^T(q^2)
    \Bigr] .
\end{align}
$T_{ij}$ is the generator of the representation for the quark.
The subscript `$q$' indicates the functions with vanishing incoming quark
momentum and `$g$' indicates those with vanishing gluon momentum.
At the tree-level we have
\begin{align}
  \Lambda_q(q^2) \bigl|_\text{tree} =
  \Lambda_g(q^2) \bigl|_\text{tree} &\,=\, 1 , \\[1mm]
  \Lambda_q^T(q^2) \bigl|_\text{tree} =
  \Lambda_g^T(q^2) \bigl|_\text{tree} &\,=\, 0 .
\label{lastdef}
\end{align}

\subsection{Renormalization}

In a generic renormalization scheme $R$, the respective renormalizations of
the gluon, ghost and quark fields can be written as
\begin{align}
  (A^B)^a_\mu &\:=\: \sqrt{Z_3^R} \: (A^R)^a_\mu , \\
  (\eta^B)^a  &\:=\: \sqrt{\tilde{Z}_3^R} \: (\eta^R)^a , \\
  \psi^B_{if} &\:=\: \sqrt{Z_2^R} \: \psi^R_{if} .
\end{align}
The superscript `$B$' indicates a bare quantity and `$R$' a renormalized 
one.
For the coupling constant,
we define $a = \als / (4\pi) = g^2 / (16\pi^2)$.
Then $a$ and the gauge parameter $\xi$ are renormalized in dimensional
regularization ($D=4-2\ep$) as follows:
\begin{align}
  a^B   &\:=\: \mu^{2\ep} Z_a^R a^R , \\[1mm]
  \xi^B &\:=\: Z_3^R \xi^R .
\end{align}
Here $\mu$ is the 't~Hooft mass scale.
We have used the fact that the gauge parameter is also renormalized by
the gluon field renormalization constant, $Z_\xi^R = Z_3^R$.
The renormalization of the self-energies and vertex functions is performed as
\begin{align}
  1 + \Pi^R &\:=\: Z_3^R (1 + \Pi^B) , \\[1mm]
  1 + \tilde{\Pi}^R &\:=\: \tilde{Z}_3^R (1 + \tilde{\Pi}^B) , \\[1mm]
  1 + \Sigma_V^R &\:=\: Z_2^R (1 + \Sigma_V^B) ,
\end{align}
and
\begin{alignat}{3}
  T_i^R &\:=\: Z_1^R T_i^B ,
    &&
    &\qquad i &= 1, 2, \\
  \tilde{\Gamma}_i^R &= \tilde{Z}_1^R \tilde{\Gamma}_i^B ,
    &&
    &\qquad i &= h, g, \\
  \Lambda_i^R &\:=\: \bar{Z}_1^R \Lambda_i^B , &\qquad
  \Lambda_i^{T,R} &\:=\: \bar{Z}_1^R \Lambda_i^{T,B} ,
    &\qquad i &= q, g ,
\end{alignat}
where the vertex renormalization constants are related to the field and coupling
renormalization constants via the Slavnov-Taylor identities by
\begin{equation}
  \sqrt{Z_a^R Z_3^R}
    \,=\, \frac{Z_1^R}{Z_3^R}
    \,=\, \frac{\tilde{Z}_1^R}{\tilde{Z}_3^R}
    \,=\, \frac{\bar{Z}_1^R}{Z_2^R} \: .
\end{equation}

In \MS{}-like schemes, the renormalization constants contain only pole terms
with respect to $\ep$ and thus take the form
\begin{equation}
  Z_i^\MS
    \:=\: 1
    + \sum_{l=1}^\infty a^l \sum_{n=1}^l \frac{Z_i^{\MS,(l,n)}}{\ep^n} \: .
\end{equation}
The coefficients $Z_i^{\MS,(l,n)}$ are determined order by order in such a way
that any renormalized Green's function becomes finite. 
The choice of the subtraction point defines the specific (\MS{}-like) 
renormalization scheme; the beta function is independent of this choice.

\section{Computations and checks}
\label{sect:compt}

The results in this article are obtained by direct computation using the 
\Forcer{} package~\cite{Ueda:2016sxw,Ueda:2016yjm,FORCER}, written in \FORM{}~%
\cite{Vermaseren:2000nd,Tentyukov:2007mu,Kuipers:2012rf}. \Forcer{} is a 
four-loop equivalent of \Mincer{}~\cite{Gorishnii:1989gt,Larin:1991fz}: 
for each topology a parametric reduction to simpler topologies is provided.
Most topologies can be reduced using direct integration of one-loop insertions,
by the triangle rule~\cite{Chetyrkin:1981qh,Tkachov:1981wb} or by the diamond rule~%
\cite{Ruijl:2015aca}. If these substructures are absent, a manually designed 
reduction scheme is required. 
Through parametric integration-by-parts (IBP) identities~\cite%
{Chetyrkin:1981qh,Tkachov:1981wb}, 
similar to the triangle rule, such reductions are found in a computer-assisted 
manner.

Great care has to be taken to prevent term blow-up from rewriting momenta to a 
new basis for the transitions between topologies. In \Mincer{} this was done 
manually, but in \Forcer{} a basis is found that minimizes the terms that are 
created automatically. 
The performance of the \Forcer{} program has been demonstrated by computing the
four-loop beta function in ten minutes in the Feynman gauge and in eight hours 
with all powers of the gauge parameter on a modern 24-core machine~%
\cite{Ueda:2016sxw,Ueda:2016yjm}.

The diagrams have been generated using Qgraf~\cite{QGRAF}. 
Subsequently, we filter propagator insertions. These insertions 
are known lower-loop propagators, and can be factorized out of the diagram,
see ref.~\cite{Herzog:2016qas}. 
Next, the topology is mapped to a built-in \Forcer{} topology, after nullifying
a leg for the vertices, and the colour factor is determined using the \FORM{}
program of ref.~\cite{vanRitbergen:1998pn}.
To extract the form factors defined above, a generalization of the projection
operators in ref.~\cite{Chetyrkin:2000dq} to a generic gauge group is used.
Then the Feynman rules are applied. The remaining Lorentz-scalar integrals 
(which include loop-momenta in numerators) are computed by the \Forcer{} program.

The computation time varied between an hour and a week, on a single computer. 
The easy cases, such as the ghost propagator and quark propagator took an hour.
The gluon propagators and ghost-ghost-gluon vertex and quark-quark-gluon vertex
took about eight hours per configuration. The triple gluon vertex was the 
hardest case and took a week per configuration on a single machine with 24 
cores. Had we chosen to compute with an expansion in $\ep$, the computations 
would have been much faster.

\pagebreak

\noindent
We have checked our setup and results in various ways:
\begin{itemize}
  \item The longitudinal component of the gluon self-energy
    $\delta^{ab} q_\mu q_\nu \Pi_L(q^2)$ was shown to be zero by an explicit
    calculation at the four-loop level.
  \item The form factor $T_3(q^2)$ of the triple-gluon vertex in
    eq.~\eqref{eq:T3} was computed and indeed vanished at the four-loop level.
  \item All the self-energies and vertex functions computed in this work were
    compared up to three loops with those in ref.~\cite{Chetyrkin:2000dq}.
    Note that the finite parts of the vertex-function results in
    ref.~\cite{Chetyrkin:2000dq} are only correct for SU($N$) gauge groups,
    since the presence of quartic Casimir operators was not taken into account
    in the reconstruction of the general case. This fact was also noted in 
    ref.~\cite{Luthe:2017ttc}.
  \item The four-loop renormalization constants and anomalous dimensions for 
    the case of SU($N$) and a general linear covariant gauge were provided in 
    ancillary files \cite{Kostja:2004anc} of ref.~\cite{Chetyrkin:2004mf}. 
    Directly after \Forcer{} was completed, we established agreement with 
    those results. For a generic group our results are in agreement with
    ancillary files of ref.~\cite{Luthe:2017ttc}.
  \item We remark that the ghost-gluon vertex is unrenormalized
    $\tilde{Z}_1^\MSbar = \tilde{Z}_3^\MSbar \sqrt{Z_a^\MSbar Z_3^\MSbar} = 1$
    in the Landau gauge.
    Moreover, our results confirm that the vertex has no radiative corrections
    when the incoming ghost momentum is nullified
    (i.e., $\tilde{\Gamma}_h^\MSbar = 1$) in the Landau gauge up to four loops.
\end{itemize}

The renormalized results for all self-energies and vertex functions are rather 
lengthy and are thus given in appendix~\ref{sec:results}\@. 
The unrenormalized results in terms of master integrals, and the values of the 
master integrals in $4-2 \ep$ dimensions are available online on
\texttt{arxiv.org}, as ancillary files to the paper. 
A description of the files is given in appendix~\ref{sec:notationfile}\@.

\section{Five-loop Landau-gauge QCD beta function in the \MiniMOM{} scheme}
\label{sect:betaMM}

In the \MiniMOM{} scheme~\cite{vonSmekal:2009ae},
the self-energies are completely absorbed into the field renormalization
constants at the subtraction point $q^2 = - \mu^2$:
\begin{align}
  1 + \Pi^\MM(-\mu^2)
    &\:=\: Z_3^\MM \Bigl[ 1 + \Pi^B(-\mu^2) \Bigr] \:=\: 1 , \\
  1 + \tilde{\Pi}^\MM(-\mu^2)
    &\:=\: \tilde{Z}_3^\MM \Bigl[ 1 + \tilde{\Pi}^B(-\mu^2) \Bigr] \:=\: 1 , \\
  1 + \Sigma_V^\MM(-\mu^2)
    &\:=\: Z_2^\MM \Bigl[ 1 + \Sigma_V^B(-\mu^2) \Bigr] \:=\: 1 .
\end{align}
Here the superscript `\MM{}' indicates a quantity in the \MiniMOM{} scheme.
In addition, motivated by the non-renormalization of the ghost-gluon
vertex in the Landau gauge~\cite{Taylor:1971ff}, the vertex renormalization
constant for this vertex is chosen the same as that in \MSbar{},
\begin{equation}
  \tilde{Z}_1^\MM \:=\: \tilde{Z}_1^\MSbar ,
\end{equation}
which is equal to one in the Landau gauge.

The above renormalization conditions lead to the following relations for
the coupling constant and gauge parameter in the two schemes:
\begin{align}
  a^\MM(\mu^2) &\:=\: a^\MSbar(\mu^2)
  \:\frac{1}{\bigl[1+\Pi^\MSbar(-\mu^2)\bigr]
           \bigl[1+\tilde{\Pi}^\MSbar(-\mu^2)\bigr]^2} \: , \label{aMM} \\[1mm]
  \xi^\MSbar(\mu^2) &\:=\: \xi^\MM(\mu^2) \:\frac{1}{1+\Pi^\MSbar(-\mu^2)} \: .
  \label{xiMM}
\end{align}
Eq.~\eqref{aMM} allows one to convert a value of $\alpha_s^\MSbar$ to 
$\alpha_s^\MM$.  For example, $\alpha_s^\MSbar(M_Z^2) = 0.118$ leads to 
$\alpha_s^\MM(M_Z^2) = 1.096 \, \alpha_s^\MSbar(M_Z^2)$ for QCD in the Landau
gauge with $n_f=5$ quark flavours.  The general expansion of eq.~\eqref{aMM} 
is given in appendix~\ref{sec:mmresults}.

The scale dependence of the coupling constant in eq.~\eqref{aMM} in this scheme
is given by
\begin{equation}
  \beta^\MM
  \:=\: \mu^2 \frac{d a^\MM}{d \mu^2}
  \:=\: \frac{\partial a^\MM}{\partial a^\MSbar} \beta^\MSbar
  \:+\: \frac{\partial a^\MM}{\partial \xi^\MSbar} \gamma_3^\MSbar \xi^\MSbar ,
  \label{betaMM}
\end{equation}
where we have used the beta function and gluon field anomalous dimension
in \MSbar{},
\begin{align}
  \beta^\MSbar &\:=\: \mu^2 \frac{d a^\MSbar}{d \mu^2} \: , \\[1mm]
  \gamma_3^\MSbar \xi^\MSbar &\:=\: \mu^2 \frac{d \xi^\MSbar}{d \mu^2} \: .
\end{align}
Note that the right-hand side of eq.~\eqref{aMM}, and hence that of
eq.~\eqref{betaMM}, is naturally given in terms of $a^\MSbar$ and $\xi^\MSbar$.
One has to convert them into $a^\MM$ and $\xi^\MM$ by inverting the series 
of eq.~\eqref{aMM} and by using eq.~\eqref{xiMM}.%
\footnote{%
  In ref.~\cite{vonSmekal:2009ae}, the results are presented in $\xi^\MSbar{}$ 
  instead of $\xi^\MM$.
  On the other hand, in ref.~\cite{Gracey:2013sca} the conversion
  from $\xi^\MSbar$ to $\xi^\MM$ was performed.
  The results become the same in the Landau gauge $\xi^\MSbar = \xi^\MM = 0$. 
  The same is true for the ``\MOMtilde{}h'' scheme of 
  ref.~\cite{Chetyrkin:2000dq}.
}

Having results for the four-loop self-energies in the \MSbar{} scheme at hand,
one can obtain the five-loop beta function in the \MiniMOM{} scheme
from the five-loop beta function \cite{Baikov:2016tgj,Herzog:2017ohr} and the 
four-loop gluon field anomalous dimension in the \MSbar{} scheme.
The result for SU(3) in the Landau gauge ($\xi^\MM=0$) reads
\begin{equation}
  \beta^{\,\MM}
    \:=\: - \sum_{l=0}^4 \bigl(a^\MM\bigr)^{l+2} \beta_l^{\,\MM}
    \:+\: \mathcal{O}\Bigl(\bigl(a^\MM\bigr)^7\Bigr) ,
\end{equation}
with
{
  \renewcommand{\colorfactor}[1]{#1}
  \def\exprname#1{\beta_{#1}^{\,\MM}}
  \def\exprfile#1{form2tex/betaLandauQCDMM#1.inc}
  \def\exprlabel#1{betaMMcoeffs#1}
  \putresultb
  \label{betaMMcoeffs}
}
The result for a generic group and in an arbitrary covariant linear gauge can
be found in appendix~\ref{sec:mmresults}; it agrees with the result given in
ref.~\cite{Gracey:2013sca} up to four loops.
As is well known, the first coefficient $\beta_0^{\,\MM}$ is scheme independent.
The next coefficient $\beta_1^{\,\MM}$ has a gauge dependence and the universal
value is obtained only in the Landau gauge.
The last coefficient $\beta_4^{\,\MM}$ is the new result.
In the \MSbar{} scheme, some of higher values of the zeta function (e.g., 
$\zeta_3^2$, $\zeta_6$ and $\zeta_7$ at five loops) do not occur, for a 
discussion of this issue see refs.~%
\cite{Baikov:2010hf,Baikov:2012zm,Baikov:2017ujl}.
In contrast, one cannot expect their absence in the \MiniMOM{} scheme.
Indeed eq.~\eqref{res:betaMMcoeffs} includes terms with $\zeta_3^2$ and 
$\zeta_7$, and for $\xi^\MM \neq 0$ also $\zeta_6$ occurs. 

The numerical values of the above beta function for three to five quark
flavours are
\begin{equation}
\label{betaMMnum}
  \begin{alignedat}{2}
    \tilde{\beta}^\MM(n_f=3) \:=\:
       1
       &+ 0.5658842421  \alpha_{\rm s}^\MM
      &&+ 0.9419859046 (\alpha_{\rm s}^\MM)^2 \\
       &+ 2.304494526  (\alpha_{\rm s}^\MM)^3
      &&+ 6.647485913  (\alpha_{\rm s}^\MM)^4 , \\
    \tilde{\beta}^\MM(n_f=4) \:=\:
       1
       &+ 0.4901972247  \alpha_{\rm s}^\MM
      &&+ 0.6452147391 (\alpha_{\rm s}^\MM)^2 \\
       &+ 1.638457168  (\alpha_{\rm s}^\MM)^3
      &&+ 3.466865543  (\alpha_{\rm s}^\MM)^4 , \\
    \tilde{\beta}^\MM(n_f=5) \:=\:
       1
       &+ 0.4013472477  \alpha_{\rm s}^\MM
      &&+ 0.3288519562 (\alpha_{\rm s}^\MM)^2 \\
       &+ 1.026892491  (\alpha_{\rm s}^\MM)^3
      &&+ 0.8417657296 (\alpha_{\rm s}^\MM)^4 ,
  \end{alignedat}
\end{equation}
where $\tilde{\beta} \equiv \beta(a)/(-\beta_0 a^2)$ has been re-expanded
in powers of $\als = 4 \pi a$.
These values may be compared with those in the \MSbar{} scheme
\cite{Baikov:2016tgj,Herzog:2017ohr} reading
\begin{equation}
\label{betaMSnum}
  \begin{alignedat}{2}
    \tilde{\beta}^\MSbar(n_f=3) \:=\:
       1
       &+ 0.5658842421  \alpha_{\rm s}^\MSbar
      &&+ 0.4530135791 (\alpha_{\rm s}^\MSbar)^2 \\
       &+ 0.6769674420 (\alpha_{\rm s}^\MSbar)^3
      &&+ 0.5809276379 (\alpha_{\rm s}^\MSbar)^4 , \\
    \tilde{\beta}^\MSbar(n_f=4) \:=\:
       1
       &+ 0.4901972247  \alpha_{\rm s}^\MSbar
      &&+ 0.3087903795 (\alpha_{\rm s}^\MSbar)^2 \\
       &+ 0.4859007965 (\alpha_{\rm s}^\MSbar)^3
      &&+ 0.2806008338 (\alpha_{\rm s}^\MSbar)^4 , \\
    \tilde{\beta}^\MSbar(n_f=5) \:=\:
       1
       &+ 0.4013472477   \alpha_{\rm s}^\MSbar
      &&+ 0.1494273313  (\alpha_{\rm s}^\MSbar)^2 \\
       &+ 0.3172233974  (\alpha_{\rm s}^\MSbar)^3
      &&+ 0.08092104151 (\alpha_{\rm s}^\MSbar)^4 .
  \end{alignedat}
\end{equation}
Obviously, the \MiniMOM{} coefficients in eqs.~(\ref{betaMMnum}) 
are (much) larger than their \MSbar{} counterparts in 
eqs.~(\ref{betaMSnum}) starting from the 
second order; moreover, they exhibit a definite growth with the
order that is absent in the \MSbar{} case. 
One may expect that this behaviour, and the larger value of 
$\alpha_{\rm s}^{\,\rm MM}$, is more than compensated by smaller expansion 
coefficients for observables, leading to a better overall convergence in 
\MOM{}-like schemes. 
However, this issue has been studied up to four loops in some detail for the 
$R$-ratio in electron-positron annihilation, without arriving at such a 
clear-cut conclusion~\cite{Gracey:2014pba}.

\section{Summary and outlook}
\label{sect:conclusion}

We have computed the four-loop three-particle vertices with one vanishing
momentum and the four-loop self-energies for QCD-like theories in a manner
that is as general as presently possible.
Our results, explicitly presented in the appendix of this article for the 
$D\!=\!4$ renormalized quantities in the \MSbar{} scheme, and available 
online in terms of their more lengthy bare counterparts, should be useful 
for precision studies of QCD-like theories 
  with any simple compact gauge group, 
  in any linear covariant gauge, 
  for any MS-like or \MOMtilde{}-like renormalization scheme 
    (see also ref.~\cite{Chetyrkin:2000dq}),
  and in any number of space-time dimensions $D$. 
The latter requires replacing the master integrals, which is the only 
component that is not known exactly in~$D$.

As an example application, we have determined the five-loop beta function in 
the \mbox{\MiniMOM{}} scheme of ref.~\cite{vonSmekal:2009ae}, i.e., we have 
extended the result of refs.~\cite{vonSmekal:2009ae,Gracey:2013sca} by one 
order in the coupling constant $\als$.
This function appears to have a higher-order structure quite different from
that in the \MSbar{} scheme, thus inviting further studies especially
for the physical case of QCD in four dimensions.

Our computations have been made possible by the construction of \Forcer{}
\cite{Ueda:2016sxw,Ueda:2016yjm,FORCER}, a four-loop generalization of the
well-known \Mincer{} program \cite{Gorishnii:1989gt,Larin:1991fz} for the
parametric reduction of three-loop massless self-energy integrals.
We have verified that, except for an issue in the transition from SU($N$) to a 
general gauge group that was also noted in ref.~\cite{Luthe:2017ttc}, our 
renormalized self-energies and vertices up to three loops agree with the 
results of ref.~\cite{Chetyrkin:2000dq}.
Furthermore, we have verified to four loops that the ghost-ghost-gluon vertex 
is unrenormalized in the Landau gauge, i.e., its anomalous dimension is 
proportional to $\xi$ and that the gluon propagator is transverse. 
Earlier, we had checked the four-loop beta function and the four-loop 
SU($N$)~\cite{Kostja:2004anc} renormalization constants and anomalous dimensions 
of ref.~\cite{Chetyrkin:2004mf}.

Performing a similar calculation at five loops is a formidable challenge: 
so far most methods used for computations at five loops use infrared 
rearrangement, which modifies the finite terms. 
A direct computation, as has been performed in this article, would require a 
five-loop \Forcer{} equivalent. This is hard for at least two reasons: the number of 
topologies that need a manually designed step-by-step IBP reduction is larger 
than 200, and the number of parameters increases from 14 at four loops to 20 
at five loops. Additionally, given the size of the step from three to four 
loops, the required computer time could be an~issue.

\acknowledgments

We would like to thank J.A.~Gracey for useful discussions.
This work has been supported by the {\it European Research Council}$\,$ (ERC)
Advanced Grant 320651, {\sc HEPGAME}, and the UK {\it Science \& Technology
Facilities Council}$\,$ (STFC) under grant number ST/L000431/1.
The figures were made with {\tt Axodraw2}~\cite{Collins:2016aya}.

\appendix

\section{Four-loop \texorpdfstring{\MSbar{}}{MSbar} results for self-energies
and vertices}
\label{sec:results}

We expand the results for the scalar `form factors' of the self-energies 
and vertices addressed in eqs.~(\ref{firstdef}) -- (\ref{lastdef}) at the 
point $q^2 = - \mu^2$ in the \MSbar{} scheme as
\begin{alignat}{2}
  \Pi^\MSbar(-\mu^2)
    &\:=\: \sum_{l=1}^4 a^l \Pi^{\MSbar,(l)} + \mathcal{O}(a^5) , \\
  \tilde{\Pi}^\MSbar(-\mu^2)
    &\:=\: \sum_{l=1}^4 a^l \tilde{\Pi}^{\MSbar,(l)} + \mathcal{O}(a^5) , \\
  \Sigma_V^\MSbar(-\mu^2)
    &\:=\: \sum_{l=1}^4 a^l \Sigma_V^{\MSbar,(l)} + \mathcal{O}(a^5) , \\
  T_1^\MSbar(-\mu^2)
    &\:=\: 1 + \sum_{l=1}^4 a^l T_1^{\MSbar,(l)} + \mathcal{O}(a^5) , \\
  T_2^\MSbar(-\mu^2)
    &\:=\: \sum_{l=1}^4 a^l T_2^{\MSbar,(l)} + \mathcal{O}(a^5) , \\
  \tilde{\Gamma}_i^\MSbar(-\mu^2)
    &\:=\: 1 + \sum_{l=1}^4 a^l \tilde{\Gamma}_i^{\MSbar,(l)}
         + \mathcal{O}(a^5) ,
    &\qquad i &= h, g, \\
  \Lambda_i^\MSbar(-\mu^2)
    &\:=\: 1 + \sum_{l=1}^4 a^l \Lambda_i^{\MSbar,(l)} + \mathcal{O}(a^5) ,
    &\qquad i &= q, g, \\
  \Lambda_i^{T,\MSbar}(-\mu^2)
    &\:=\: \sum_{l=1}^4 a^l \Lambda_i^{T,\MSbar,(l)} + \mathcal{O}(a^5) ,
    &\qquad i &= q, g.
\end{alignat}
Here the coupling constant $a = \alpha_s/(4\pi) = g^2/(16\pi^2)$ and
the gauge parameter $\xi$ are the renormalized ones in the \MSbar{} scheme, 
i.e., $a^\MSbar(\mu^2)$ and $\xi^\MSbar(\mu^2)$.
The Landau gauge corresponds to $\xi = 0$. Recall that this differs from
the convention in \Mincer{} and \Forcer{}, where $\xi = 0$ corresponds to
the Feynman gauge and $\xi = 1$ to the Landau gauge.

\subsection{Gluon self-energy}
\def\exprname#1{\Pi^{\MSbar,(#1)}}
\def\exprfile#1{form2tex/gluonMSb#1.inc}

    \begin{align}
      \begin{autobreak}
        \MoveEqLeft
        \exprname{1} =
        \biggl(\frac{34}{3} \colorfactor{\ca^2}
-\frac{20}{3} \colorfactor{\ca \tf \nf}
-4 \colorfactor{\cf \tf \nf}\biggr)
+\biggl(-\frac{13}{12} \colorfactor{\ca^2}
+\frac{2}{3} \colorfactor{\ca \tf \nf}\biggr) \gaugefactor{\xil}
+\biggl(-\frac{5}{6} \colorfactor{\ca^2}
+\frac{2}{3} \colorfactor{\ca \tf \nf}\biggr) \gaugefactor{\xil^2}
+\frac{1}{4} \colorfactor{\ca^2} \gaugefactor{\xil^3}
 ,
      \end{autobreak} \nonumber \\[1mm]
      \begin{autobreak}
        \MoveEqLeft
        \exprname{2} =
        \Biggl[\colorfactor{\ca^3} \biggl(\frac{9655}{72}
-\frac{143}{8} \z3\biggr)
+\colorfactor{\ca^2 \tf \nf} \biggl(-\frac{2009}{18}
-\frac{137}{6} \z3\biggr)
+\colorfactor{\ca \cf \tf \nf} \biggl(-\frac{641}{9}
+\frac{176}{3} \z3\biggr)
+2 \colorfactor{\cf^2 \tf \nf}
+\colorfactor{\ca \tf^2 \nf^2} \biggl(\frac{46}{3}
+\frac{32}{3} \z3\biggr)
+\colorfactor{\cf \tf^2 \nf^2} \biggl(\frac{184}{9}
-\frac{64}{3} \z3\biggr)\Biggr]
+\Biggl[\colorfactor{\ca^3} \biggl(-\frac{1097}{96}
+\frac{33}{8} \z3\biggr)
+\frac{37}{6} \colorfactor{\ca^2 \tf \nf}
+4 \colorfactor{\ca \cf \tf \nf}\Biggr] \gaugefactor{\xil}
+\Biggl[\colorfactor{\ca^3} \biggl(-\frac{725}{96}
+\frac{13}{8} \z3\biggr)
+\colorfactor{\ca^2 \tf \nf} \biggl(\frac{23}{4}
-\frac{1}{2} \z3\biggr)
+3 \colorfactor{\ca \cf \tf \nf}\Biggr] \gaugefactor{\xil^2}
+\Biggl[\colorfactor{\ca^3} \biggl(\frac{21}{32}
-\frac{3}{8} \z3\biggr)
+\frac{1}{2} \colorfactor{\ca^2 \tf \nf}\Biggr] \gaugefactor{\xil^3}
+\biggl(\frac{55}{96} \colorfactor{\ca^3}
-\frac{1}{12} \colorfactor{\ca^2 \tf \nf}\biggr) \gaugefactor{\xil^4}
 ,
      \end{autobreak} \nonumber \\[1mm]
      \begin{autobreak}
        \MoveEqLeft
        \exprname{3} =
        \Biggl[\colorfactor{\ca^4} \biggl(\frac{1381669}{648}
-\frac{229559}{432} \z3
-\frac{85855}{288} \z5\biggr)
+\colorfactor{\dfAAna} \biggl(-\frac{80}{9}
+\frac{704}{3} \z3\biggr)
+\colorfactor{\ca^3 \tf \nf} \biggl(-\frac{244805}{108}
+\frac{1409}{12} \z3
+\frac{35965}{72} \z5\biggr)
+\colorfactor{\ca^2 \cf \tf \nf} \biggl(-\frac{60685}{36}
+680 \z3
+\frac{1760}{3} \z5\biggr)
+\colorfactor{\ca \cf^2 \tf \nf} \biggl(\frac{527}{9}
+\frac{2288}{3} \z3
-\frac{3520}{3} \z5\biggr)
+46 \colorfactor{\cf^3 \tf \nf}
+\colorfactor{\dfFAna \nf} \biggl(\frac{512}{9}
-\frac{1664}{3} \z3\biggr)
+\colorfactor{\ca^2 \tf^2 \nf^2} \biggl(\frac{1655}{3}
+\frac{1436}{9} \z3
-\frac{1280}{9} \z5\biggr)
+\colorfactor{\ca \cf \tf^2 \nf^2} \biggl(\frac{9428}{9}
-\frac{1376}{3} \z3
-\frac{640}{3} \z5\biggr)
+\colorfactor{\cf^2 \tf^2 \nf^2} \biggl(-\frac{232}{9}
-\frac{1024}{3} \z3
+\frac{1280}{3} \z5\biggr)
+\colorfactor{\dfFFna \nf^2} \biggl(-\frac{704}{9}
+\frac{512}{3} \z3\biggr)
+\colorfactor{\ca \tf^3 \nf^3} \biggl(-\frac{1792}{81}
-\frac{896}{27} \z3\biggr)
+\colorfactor{\cf \tf^3 \nf^3} \biggl(-128
+\frac{256}{3} \z3\biggr)\Biggr]
+\Biggl[\colorfactor{\ca^4} \biggl(-\frac{191047}{1152}
+\frac{200059}{864} \z3
-\frac{9145}{96} \z5\biggr)
+\colorfactor{\ca^3 \tf \nf} \biggl(\frac{75745}{432}
-\frac{5749}{36} \z3
+\frac{295}{12} \z5\biggr)
+\colorfactor{\ca^2 \cf \tf \nf} \biggl(\frac{1193}{12}
-77 \z3\biggr)
-3 \colorfactor{\ca \cf^2 \tf \nf}
+\colorfactor{\ca^2 \tf^2 \nf^2} \biggl(-\frac{2039}{54}
+\frac{536}{9} \z3\biggr)
+\colorfactor{\ca \cf \tf^2 \nf^2} \biggl(-\frac{92}{3}
+32 \z3\biggr)
+\colorfactor{\ca \tf^3 \nf^3} \biggl(\frac{64}{27}
-\frac{256}{27} \z3\biggr)\Biggr] \gaugefactor{\xil}
+\Biggl[\colorfactor{\ca^4} \biggl(-\frac{57161}{576}
+\frac{5519}{144} \z3
-\frac{235}{16} \z5\biggr)
+\colorfactor{\ca^3 \tf \nf} \biggl(\frac{15263}{144}
+\frac{127}{36} \z3\biggr)
+\colorfactor{\ca^2 \cf \tf \nf} \biggl(\frac{686}{9}
-\frac{185}{3} \z3\biggr)
-2 \colorfactor{\ca \cf^2 \tf \nf}
+\colorfactor{\ca^2 \tf^2 \nf^2} \biggl(-\frac{116}{9}
-\frac{64}{9} \z3\biggr)
+\colorfactor{\ca \cf \tf^2 \nf^2} \biggl(-\frac{184}{9}
+\frac{64}{3} \z3\biggr)\Biggr] \gaugefactor{\xil^2}
+\Biggl[\colorfactor{\ca^4} \biggl(\frac{9097}{1152}
-\frac{1307}{288} \z3
-\frac{275}{144} \z5\biggr)
+\colorfactor{\ca^3 \tf \nf} \biggl(\frac{941}{144}
-\frac{16}{9} \z3
-\frac{5}{36} \z5\biggr)
+\frac{3}{4} \colorfactor{\ca^2 \cf \tf \nf}\Biggr] \gaugefactor{\xil^3}
+\Biggl[\colorfactor{\ca^4} \biggl(\frac{9061}{1152}
-\frac{451}{144} \z3
-\frac{115}{288} \z5\biggr)
+\colorfactor{\ca^3 \tf \nf} \biggl(-\frac{25}{36}
+\frac{5}{36} \z3
+\frac{35}{72} \z5\biggr)
-\frac{3}{4} \colorfactor{\ca^2 \cf \tf \nf}\Biggr] \gaugefactor{\xil^4}
+\Biggl[\colorfactor{\ca^4} \biggl(\frac{145}{128}
-\frac{1}{12} \z3
+\frac{35}{96} \z5\biggr)
-\frac{1}{6} \colorfactor{\ca^3 \tf \nf}\Biggr] \gaugefactor{\xil^5}
 ,
      \end{autobreak} \nonumber \\[1mm]
      \begin{autobreak}
        \MoveEqLeft
        \exprname{4} =
        \Biggl[\colorfactor{\ca^5} \biggl(\frac{361398305}{6912}
-\frac{82264805}{6912} \z3
-\frac{349180015}{9216} \z5
+\frac{7423031}{4608} \z3^2
+\frac{31354477}{1536} \z7\biggr)
+\colorfactor{\ca \dfAAna} \biggl(-\frac{72745}{216}
+\frac{9257251}{576} \z3
-\frac{3094355}{384} \z5
-\frac{1940675}{192} \z3^2
+\frac{3214211}{512} \z7\biggr)
+\colorfactor{\ca^4 \tf \nf} \biggl(-\frac{370399597}{5184}
+\frac{13787695}{1728} \z3
+\frac{372847895}{6912} \z5
-\frac{249463}{384} \z3^2
-\frac{13317479}{384} \z7\biggr)
+\colorfactor{\ca^3 \cf \tf \nf} \biggl(-\frac{20916689}{432}
+\frac{142513}{36} \z3
+\frac{1777405}{48} \z5
-3234 \z3^2
+\frac{30800}{3} \z7\biggr)
+\colorfactor{\ca^2 \cf^2 \tf \nf} \biggl(\frac{789839}{216}
+\frac{1105007}{36} \z3
-\frac{58400}{3} \z5
-24640 \z7\biggr)
+\colorfactor{\ca \cf^3 \tf \nf} \biggl(\frac{19097}{6}
+\frac{2264}{3} \z3
-\frac{72160}{3} \z5
+24640 \z7\biggr)
+\colorfactor{\cf^4 \tf \nf} \biggl(-\frac{4157}{6}
-128 \z3\biggr)
+\colorfactor{\ca \dfFAna \nf} \biggl(\frac{62288}{27}
-\frac{104944}{3} \z3
+\frac{33860}{9} \z5
+\frac{59840}{3} \z3^2
-\frac{539}{2} \z7\biggr)
+\colorfactor{\cf \dfFAna \nf} \biggl(-320
+\frac{1280}{3} \z3
+\frac{6400}{3} \z5\biggr)
+\colorfactor{\dfAAna \tf \nf} \biggl(\frac{7931}{54}
-\frac{264899}{48} \z3
+\frac{434315}{288} \z5
+\frac{176425}{48} \z3^2
-\frac{292201}{128} \z7\biggr)
+\colorfactor{\ca^3 \tf^2 \nf^2} \biggl(\frac{6019289}{216}
+\frac{19145}{4} \z3
-\frac{59675}{3} \z5
+\frac{1243}{3} \z3^2
+9912 \z7\biggr)
+\colorfactor{\ca^2 \cf \tf^2 \nf^2} \biggl(\frac{829223}{18}
-\frac{50771}{9} \z3
-\frac{294560}{9} \z5
+\frac{712}{3} \z3^2
-\frac{11200}{3} \z7\biggr)
+\colorfactor{\ca \cf^2 \tf^2 \nf^2} \biggl(-\frac{70298}{27}
-23440 \z3
+\frac{56960}{3} \z5
+\frac{2816}{3} \z3^2
+8960 \z7\biggr)
+\colorfactor{\cf^3 \tf^2 \nf^2} \biggl(-\frac{1148}{3}
-\frac{2176}{3} \z3
+\frac{30080}{3} \z5
-8960 \z7\biggr)
+\colorfactor{\ca \dfFFna \nf^2} \biggl(-\frac{110768}{27}
+\frac{51968}{9} \z3
-\frac{29440}{9} \z5
+\frac{5632}{3} \z3^2\biggr)
+\colorfactor{\cf \dfFFna \nf^2} \biggl(\frac{4160}{3}
+\frac{5120}{3} \z3
-\frac{12800}{3} \z5\biggr)
+\colorfactor{\dfFAna \tf \nf^2} \biggl(-\frac{18976}{27}
+\frac{100480}{9} \z3
+\frac{10960}{9} \z5
-\frac{21760}{3} \z3^2
+98 \z7\biggr)
+\colorfactor{\ca^2 \tf^3 \nf^3} \biggl(-\frac{94906}{27}
-\frac{74396}{27} \z3
+\frac{6220}{3} \z5
-\frac{1280}{9} \z3^2\biggr)
+\colorfactor{\ca \cf \tf^3 \nf^3} \biggl(-\frac{330676}{27}
+\frac{25888}{9} \z3
+\frac{74240}{9} \z5
+\frac{1024}{3} \z3^2\biggr)
+\colorfactor{\cf^2 \tf^3 \nf^3} \biggl(\frac{1000}{9}
+\frac{35968}{9} \z3
-\frac{12800}{3} \z5
-\frac{1024}{3} \z3^2\biggr)
+\colorfactor{\dfFFna \tf \nf^3} \biggl(\frac{24064}{27}
-\frac{9728}{9} \z3
+\frac{5120}{3} \z5
-\frac{2048}{3} \z3^2\biggr)
+\colorfactor{\ca \tf^4 \nf^4} \biggl(\frac{7568}{81}
+\frac{6400}{27} \z3
-\frac{2560}{27} \z5\biggr)
+\colorfactor{\cf \tf^4 \nf^4} \biggl(\frac{8576}{9}
-\frac{3584}{9} \z3
-\frac{5120}{9} \z5\biggr)\Biggr]
+\Biggl[\colorfactor{\ca^5} \biggl(-\frac{7426894751}{2239488}
+\frac{5670451}{1536} \z3
-\frac{55067}{1536} \z4
-\frac{92929997}{55296} \z5
+\frac{192125}{3072} \z6
-\frac{54149}{1024} \z3^2
+\frac{380293921}{221184} \z7\biggr)
+\colorfactor{\ca \dfAAna} \biggl(\frac{1217}{96}
+\frac{1160503}{1152} \z3
-\frac{5883}{64} \z4
-\frac{7311565}{2304} \z5
+\frac{1325}{8} \z6
+\frac{138171}{128} \z3^2
-\frac{393631}{2304} \z7\biggr)
+\colorfactor{\ca^4 \tf \nf} \biggl(\frac{336999223}{93312}
-\frac{1678727}{432} \z3
+\frac{4733}{48} \z4
+\frac{1188881}{1152} \z5
-\frac{3625}{192} \z6
+\frac{64553}{576} \z3^2
-\frac{15402653}{13824} \z7\biggr)
+\colorfactor{\ca^3 \cf \tf \nf} \biggl(\frac{817505}{216}
-\frac{33455}{18} \z3
-106 \z4
-\frac{4421}{3} \z5
-\frac{968}{3} \z3^2\biggr)
+\colorfactor{\ca^2 \cf^2 \tf \nf} \biggl(-\frac{994}{9}
-\frac{4609}{3} \z3
+\frac{7040}{3} \z5\biggr)
-92 \colorfactor{\ca \cf^3 \tf \nf}
+\colorfactor{\ca \dfFAna \nf} \biggl(-\frac{1024}{9}
-\frac{16736}{9} \z3
+\frac{9190}{3} \z5
-\frac{424}{3} \z3^2
+\frac{2597}{4} \z7\biggr)
+\colorfactor{\dfAAna \tf \nf} \biggl(\frac{8}{9}
-\frac{30443}{72} \z3
+\frac{111}{4} \z4
+\frac{134495}{144} \z5
-50 \z6
-\frac{2607}{8} \z3^2
+\frac{7427}{144} \z7\biggr)
+\colorfactor{\ca^3 \tf^2 \nf^2} \biggl(-\frac{233570}{243}
+\frac{26189}{18} \z3
-\frac{265}{6} \z4
-\frac{7726}{27} \z5
-\frac{388}{9} \z3^2
+\frac{539}{3} \z7\biggr)
+\colorfactor{\ca^2 \cf \tf^2 \nf^2} \biggl(-\frac{63032}{27}
+\frac{4592}{3} \z3
+32 \z4
+\frac{1552}{3} \z5
+\frac{352}{3} \z3^2\biggr)
+\colorfactor{\ca \cf^2 \tf^2 \nf^2} \biggl(\frac{464}{9}
+\frac{2048}{3} \z3
-\frac{2560}{3} \z5\biggr)
+\colorfactor{\ca \dfFFna \nf^2} \biggl(\frac{1408}{9}
-\frac{1024}{3} \z3\biggr)
+\colorfactor{\dfFAna \tf \nf^2} \biggl(\frac{7936}{9} \z3
-\frac{2720}{3} \z5
+\frac{128}{3} \z3^2
-196 \z7\biggr)
+\colorfactor{\ca^2 \tf^3 \nf^3} \biggl(\frac{121048}{2187}
-\frac{5320}{27} \z3
+\frac{16}{3} \z4
+\frac{1360}{27} \z5\biggr)
+\colorfactor{\ca \cf \tf^3 \nf^3} \biggl(\frac{2624}{9}
-\frac{2816}{9} \z3\biggr)\Biggr] \gaugefactor{\xil}
+\Biggl[\colorfactor{\ca^5} \biggl(-\frac{1477213633}{746496}
+\frac{10039865}{13824} \z3
-\frac{10177}{1536} \z4
+\frac{7102253}{55296} \z5
+\frac{26875}{3072} \z6
-\frac{104137}{3072} \z3^2
+\frac{3658879}{24576} \z7\biggr)
+\colorfactor{\ca \dfAAna} \biggl(\frac{2983}{288}
-\frac{149371}{1152} \z3
-\frac{993}{64} \z4
-\frac{113705}{768} \z5
+\frac{275}{16} \z6
+\frac{85543}{384} \z3^2
-\frac{164647}{768} \z7\biggr)
+\colorfactor{\ca^4 \tf \nf} \biggl(\frac{224764001}{93312}
-\frac{1727}{18} \z3
+\frac{3061}{384} \z4
-\frac{1885135}{3456} \z5
-\frac{25}{24} \z6
-\frac{23957}{576} \z3^2
-\frac{50687}{1152} \z7\biggr)
+\colorfactor{\ca^3 \cf \tf \nf} \biggl(\frac{321707}{144}
-\frac{6619}{6} \z3
-6 \z4
-\frac{6061}{8} \z5
+66 \z3^2\biggr)
+\colorfactor{\ca^2 \cf^2 \tf \nf} \biggl(-\frac{5513}{72}
-\frac{11413}{12} \z3
+\frac{4400}{3} \z5\biggr)
-\frac{115}{2} \colorfactor{\ca \cf^3 \tf \nf}
+\colorfactor{\ca \dfFAna \nf} \biggl(-\frac{640}{9}
+480 \z3
+230 \z5
-8 \z3^2
+\frac{147}{4} \z7\biggr)
+\colorfactor{\dfAAna \tf \nf} \biggl(-\frac{7}{18}
+\frac{83}{72} \z3
+\frac{33}{16} \z4
-\frac{365}{16} \z5
-\frac{25}{16} \z6
-\frac{97}{3} \z3^2
+\frac{7861}{192} \z7\biggr)
+\colorfactor{\ca^3 \tf^2 \nf^2} \biggl(-\frac{6644311}{11664}
-\frac{27881}{108} \z3
-\frac{17}{12} \z4
+\frac{19475}{108} \z5
+\frac{133}{9} \z3^2\biggr)
+\colorfactor{\ca^2 \cf \tf^2 \nf^2} \biggl(-\frac{23861}{18}
+659 \z3
+\frac{800}{3} \z5
-24 \z3^2\biggr)
+\colorfactor{\ca \cf^2 \tf^2 \nf^2} \biggl(\frac{290}{9}
+\frac{1280}{3} \z3
-\frac{1600}{3} \z5\biggr)
+\colorfactor{\ca \dfFFna \nf^2} \biggl(\frac{880}{9}
-\frac{640}{3} \z3\biggr)
+\colorfactor{\ca^2 \tf^3 \nf^3} \biggl(\frac{1516}{81}
+\frac{1400}{27} \z3
-\frac{40}{27} \z5\biggr)
+\colorfactor{\ca \cf \tf^3 \nf^3} \biggl(160
-\frac{320}{3} \z3\biggr)\Biggr] \gaugefactor{\xil^2}
+\Biggl[\colorfactor{\ca^5} \biggl(\frac{5211575}{27648}
-\frac{1711553}{6912} \z3
-\frac{785}{512} \z4
+\frac{206809}{3072} \z5
+\frac{1925}{1024} \z6
+\frac{385}{64} \z3^2
+\frac{278971}{12288} \z7\biggr)
+\colorfactor{\ca \dfAAna} \biggl(-\frac{17}{96}
-\frac{6877}{192} \z3
-\frac{423}{64} \z4
+\frac{12835}{128} \z5
+\frac{375}{32} \z6
+\frac{2263}{64} \z3^2
-\frac{1505}{32} \z7\biggr)
+\colorfactor{\ca^4 \tf \nf} \biggl(-\frac{15245}{864}
+\frac{2839}{16} \z3
+\frac{5}{16} \z4
-\frac{9625}{288} \z5
-\frac{25}{12} \z3^2\biggr)
+\colorfactor{\ca^3 \cf \tf \nf} \biggl(\frac{1969}{144}
-\frac{9}{2} \z3
-\frac{5}{6} \z5\biggr)
+\frac{1}{2} \colorfactor{\ca^2 \cf^2 \tf \nf}
+\colorfactor{\ca^3 \tf^2 \nf^2} \biggl(\frac{1763}{108}
-\frac{218}{3} \z3
+\frac{10}{9} \z5\biggr)
+\colorfactor{\ca^2 \cf \tf^2 \nf^2} \biggl(\frac{46}{9}
-\frac{16}{3} \z3\biggr)
+\colorfactor{\ca^2 \tf^3 \nf^3} \biggl(-\frac{64}{27}
+\frac{256}{27} \z3\biggr)\Biggr] \gaugefactor{\xil^3}
+\Biggl[\colorfactor{\ca^5} \biggl(\frac{2993563}{27648}
-\frac{221371}{3456} \z3
-\frac{439}{1536} \z4
+\frac{194767}{13824} \z5
+\frac{2675}{9216} \z6
+\frac{11947}{4608} \z3^2
+\frac{69545}{12288} \z7\biggr)
+\colorfactor{\ca \dfAAna} \biggl(-\frac{3}{32}
-\frac{1987}{288} \z3
-\frac{59}{64} \z4
+\frac{3485}{144} \z5
+\frac{125}{48} \z6
-\frac{17}{24} \z3^2
+\frac{1379}{512} \z7\biggr)
+\colorfactor{\ca^4 \tf \nf} \biggl(-\frac{21649}{864}
+\frac{9943}{1728} \z3
-\frac{1}{384} \z4
-\frac{11545}{6912} \z5
-\frac{25}{576} \z6
-\frac{965}{1152} \z3^2
+\frac{49}{128} \z7\biggr)
+\colorfactor{\ca^3 \cf \tf \nf} \biggl(-\frac{1351}{48}
+\frac{95}{4} \z3
+\frac{35}{16} \z5\biggr)
+\frac{3}{4} \colorfactor{\ca^2 \cf^2 \tf \nf}
+\colorfactor{\dfAAna \tf \nf} \biggl(\frac{691}{144} \z3
+\frac{7}{16} \z4
-\frac{1385}{288} \z5
-\frac{25}{48} \z6
-\frac{121}{48} \z3^2
+\frac{595}{128} \z7\biggr)
+\colorfactor{\ca^3 \tf^2 \nf^2} \biggl(\frac{47}{12}
+\frac{4}{3} \z3\biggr)
+\colorfactor{\ca^2 \cf \tf^2 \nf^2} \biggl(\frac{23}{3}
-8 \z3\biggr)\Biggr] \gaugefactor{\xil^4}
+\Biggl[\colorfactor{\ca^5} \biggl(\frac{60875}{2304}
-\frac{36863}{4608} \z3
-\frac{1}{256} \z4
+\frac{19355}{55296} \z5
-\frac{25}{384} \z6
-\frac{1865}{3072} \z3^2
+\frac{122185}{221184} \z7\biggr)
+\colorfactor{\ca \dfAAna} \biggl(\frac{935}{128} \z3
+\frac{21}{32} \z4
-\frac{14585}{2304} \z5
-\frac{25}{32} \z6
-\frac{489}{128} \z3^2
+\frac{16247}{2304} \z7\biggr)
+\colorfactor{\ca^4 \tf \nf} \biggl(-\frac{103}{32}
+\frac{65}{48} \z3
+\frac{7325}{3456} \z5
-\frac{7}{64} \z3^2
-\frac{4823}{13824} \z7\biggr)
-\frac{9}{8} \colorfactor{\ca^3 \cf \tf \nf}
+\colorfactor{\dfAAna \tf \nf} \biggl(\frac{5}{24} \z3
-\frac{35}{144} \z5
-\frac{5}{8} \z3^2
+\frac{91}{72} \z7\biggr)\Biggr] \gaugefactor{\xil^5}
+\Biggl[\colorfactor{\ca^5} \biggl(\frac{2015}{1152}
+\frac{1523}{4608} \z3
+\frac{39275}{18432} \z5
-\frac{105}{1024} \z3^2
-\frac{24115}{73728} \z7\biggr)
+\colorfactor{\ca \dfAAna} \biggl(\frac{33}{128} \z3
-\frac{145}{768} \z5
-\frac{75}{128} \z3^2
+\frac{455}{384} \z7\biggr)
+\colorfactor{\ca^4 \tf \nf} \biggl(-\frac{61}{288}
-\frac{1}{288} \z3
-\frac{35}{288} \z5\biggr)
+\frac{1}{16} \colorfactor{\ca^3 \cf \tf \nf}\Biggr] \gaugefactor{\xil^6}
 .
      \end{autobreak}
    \end{align}

\subsection{Ghost self-energy}
\def\exprname#1{\tilde{\Pi}^{\MSbar,(#1)}}
\def\exprfile#1{form2tex/ghostMSb#1.inc}

    \begin{align}
      \begin{autobreak}
        \MoveEqLeft
        \exprname{1} =
         ,
      \end{autobreak} \nonumber \\[1mm]
      \begin{autobreak}
        \MoveEqLeft
        \exprname{2} =
         ,
      \end{autobreak} \nonumber \\[1mm]
      \begin{autobreak}
        \MoveEqLeft
        \exprname{3} =
         ,
      \end{autobreak} \nonumber \\[1mm]
      \begin{autobreak}
        \MoveEqLeft
        \exprname{4} =
         .
      \end{autobreak}
    \end{align}

\subsection{Quark self-energy}
\def\exprname#1{\Sigma_V^{\MSbar,(#1)}}
\def\exprfile#1{form2tex/quarkMSb#1.inc}

    \begin{align}
      \begin{autobreak}
        \MoveEqLeft
        \exprname{1} =
         ,
      \end{autobreak} \nonumber \\[1mm]
      \begin{autobreak}
        \MoveEqLeft
        \exprname{2} =
         ,
      \end{autobreak} \nonumber \\[1mm]
      \begin{autobreak}
        \MoveEqLeft
        \exprname{3} =
         ,
      \end{autobreak} \nonumber \\[1mm]
      \begin{autobreak}
        \MoveEqLeft
        \exprname{4} =
         .
      \end{autobreak}
    \end{align}

\subsection{Triple-gluon vertex}
\def\exprname#1{T_1^{\MSbar,(#1)}}
\def\exprfile#1{form2tex/glglgl1MSb#1.inc}

    \begin{align}
      \begin{autobreak}
        \MoveEqLeft
        \exprname{1} =
         ,
      \end{autobreak} \nonumber \\[1mm]
      \begin{autobreak}
        \MoveEqLeft
        \exprname{2} =
         ,
      \end{autobreak} \nonumber \\[1mm]
      \begin{autobreak}
        \MoveEqLeft
        \exprname{3} =
         ,
      \end{autobreak} \nonumber \\[1mm]
      \begin{autobreak}
        \MoveEqLeft
        \exprname{4} =
         .
      \end{autobreak}
    \end{align}
  
\def\exprname#1{T_2^{\MSbar,(#1)}}
\def\exprfile#1{form2tex/glglgl2MSb#1.inc}

    \begin{align}
      \begin{autobreak}
        \MoveEqLeft
        \exprname{1} =
         ,
      \end{autobreak} \nonumber \\[1mm]
      \begin{autobreak}
        \MoveEqLeft
        \exprname{2} =
         ,
      \end{autobreak} \nonumber \\[1mm]
      \begin{autobreak}
        \MoveEqLeft
        \exprname{3} =
         ,
      \end{autobreak} \nonumber \\[1mm]
      \begin{autobreak}
        \MoveEqLeft
        \exprname{4} =
         .
      \end{autobreak}
    \end{align}

\subsection{Ghost-gluon vertex}
\def\exprname#1{\tilde{\Gamma}_h^{\MSbar,(#1)}}
\def\exprfile#1{form2tex/ghghglhMSb#1.inc}

    \begin{align}
      \begin{autobreak}
        \MoveEqLeft
        \exprname{1} =
         ,
      \end{autobreak} \nonumber \\[1mm]
      \begin{autobreak}
        \MoveEqLeft
        \exprname{2} =
         ,
      \end{autobreak} \nonumber \\[1mm]
      \begin{autobreak}
        \MoveEqLeft
        \exprname{3} =
         ,
      \end{autobreak} \nonumber \\[1mm]
      \begin{autobreak}
        \MoveEqLeft
        \exprname{4} =
         .
      \end{autobreak}
    \end{align}
  
\def\exprname#1{\tilde{\Gamma}_g^{\MSbar,(#1)}}
\def\exprfile#1{form2tex/ghghglgMSb#1.inc}

    \begin{align}
      \begin{autobreak}
        \MoveEqLeft
        \exprname{1} =
         ,
      \end{autobreak} \nonumber \\[1mm]
      \begin{autobreak}
        \MoveEqLeft
        \exprname{2} =
         ,
      \end{autobreak} \nonumber \\[1mm]
      \begin{autobreak}
        \MoveEqLeft
        \exprname{3} =
         ,
      \end{autobreak} \nonumber \\[1mm]
      \begin{autobreak}
        \MoveEqLeft
        \exprname{4} =
         .
      \end{autobreak}
    \end{align}

\subsection{Quark-gluon vertex}
\def\exprname#1{\Lambda_q^{\MSbar,(#1)}}
\def\exprfile#1{form2tex/ququglqMSb#1.inc}

    \begin{align}
      \begin{autobreak}
        \MoveEqLeft
        \exprname{1} =
         ,
      \end{autobreak} \nonumber \\[1mm]
      \begin{autobreak}
        \MoveEqLeft
        \exprname{2} =
         ,
      \end{autobreak} \nonumber \\[1mm]
      \begin{autobreak}
        \MoveEqLeft
        \exprname{3} =
         ,
      \end{autobreak} \nonumber \\[1mm]
      \begin{autobreak}
        \MoveEqLeft
        \exprname{4} =
         .
      \end{autobreak}
    \end{align}
  
\def\exprname#1{\Lambda_q^{T,\MSbar,(#1)}}
\def\exprfile#1{form2tex/ququglTqMSb#1.inc}

    \begin{align}
      \begin{autobreak}
        \MoveEqLeft
        \exprname{1} =
         ,
      \end{autobreak} \nonumber \\[1mm]
      \begin{autobreak}
        \MoveEqLeft
        \exprname{2} =
         ,
      \end{autobreak} \nonumber \\[1mm]
      \begin{autobreak}
        \MoveEqLeft
        \exprname{3} =
         ,
      \end{autobreak} \nonumber \\[1mm]
      \begin{autobreak}
        \MoveEqLeft
        \exprname{4} =
         .
      \end{autobreak}
    \end{align}
  
\def\exprname#1{\Lambda_g^{\MSbar,(#1)}}
\def\exprfile#1{form2tex/ququglgMSb#1.inc}

    \begin{align}
      \begin{autobreak}
        \MoveEqLeft
        \exprname{1} =
         ,
      \end{autobreak} \nonumber \\[1mm]
      \begin{autobreak}
        \MoveEqLeft
        \exprname{2} =
         ,
      \end{autobreak} \nonumber \\[1mm]
      \begin{autobreak}
        \MoveEqLeft
        \exprname{3} =
         ,
      \end{autobreak} \nonumber \\[1mm]
      \begin{autobreak}
        \MoveEqLeft
        \exprname{4} =
         .
      \end{autobreak}
    \end{align}
  
\def\exprname#1{\Lambda_g^{T,\MSbar,(#1)}}
\def\exprfile#1{form2tex/ququglTgMSb#1.inc}

    \begin{align}
      \begin{autobreak}
        \MoveEqLeft
        \exprname{1} =
         ,
      \end{autobreak} \nonumber \\[1mm]
      \begin{autobreak}
        \MoveEqLeft
        \exprname{2} =
         ,
      \end{autobreak} \nonumber \\[1mm]
      \begin{autobreak}
        \MoveEqLeft
        \exprname{3} =
         ,
      \end{autobreak} \nonumber \\[1mm]
      \begin{autobreak}
        \MoveEqLeft
        \exprname{4} =
         .
      \end{autobreak}
    \end{align}

\section{General five-loop \MiniMOM{} coupling constant and beta function}
\label{sec:mmresults}

The right-hand side of eq.~\eqref{aMM} is expanded as
\begin{equation}
  \frac{a^\MM(\mu^2)}{a^\MSbar(\mu^2)}
    = 1 + \sum_{l=1}^4 a^l c^{(l)} + \mathcal{O}(a^5) 
\end{equation}
where $a$ and $\xi$ are understood as $a^\MSbar(\mu^2)$ and 
$\xi^\MSbar(\mu^2)$. The coefficients $c^{(l)}$ are given by
\def\exprname#1{c^{(#1)}}
\def\exprfile#1{form2tex/aMM#1.inc}

    \begin{align}
      \begin{autobreak}
        \MoveEqLeft
        \exprname{1} =
         ,
      \end{autobreak} \nonumber \\[1mm]
      \begin{autobreak}
        \MoveEqLeft
        \exprname{2} =
         ,
      \end{autobreak} \nonumber \\[1mm]
      \begin{autobreak}
        \MoveEqLeft
        \exprname{3} =
         ,
      \end{autobreak} \nonumber \\[1mm]
      \begin{autobreak}
        \MoveEqLeft
        \exprname{4} =
         .
      \end{autobreak}
    \end{align}

\noindent
In terms of $a^\MM(\mu^2)$ and $\xi^\MM(\mu^2)$, 
the beta function in eq.~\eqref{betaMM} is expanded as
\begin{equation}
  \beta^\MM(a^\MM)
  = \mu^2 \frac{d a^\MM(\mu^2)}{d \mu^2}
    = - \sum_{l=0}^4 a^{l+2} \beta_l^\MM + \mathcal{O}(a^7) 
\end{equation}
with

\vspace*{-4mm}
\def\exprname#1{\beta_{#1}^\MM}
\def\exprfile#1{form2tex/betaMM#1.inc}
\putresultb

\section{Notations in the result files}
\label{sec:notationfile}

The notation of the result file, which is distributed along with this article 
on \texttt{arxiv.org}, is displayed in three tables below. 
We define the colour factors in table~\ref{tab:colour} (see ref.~%
\cite{vanRitbergen:1998pn} for the definitions), and the names of the 
unrenormalized and renormalized results in table~\ref{tab:unrenorm} and
table~\ref{tab:renorm} respectively. We also provide transition functions 
from the calculationally convenient G-scheme \cite{Chetyrkin:1980pr}, where a 
certain universal function is set to one, to the standard convention used 
for calculations in \MSbar, as well as values for the master integrals in 
$4-2\ep$ dimensions, which were adapted for use by \Forcer{} from the results
in refs.~\cite{Baikov:2010hf,Lee:2011jt}.

\begin{table}[hb]
  \centering
  \renewcommand{\arraystretch}{1.2}
  \begin{tabular}{MTMMM}
    \hline
                                     &           & \text{SU($N$)}                     & \text{QCD}    & \text{QED} \\
    \hline \rule{0pt}{4ex}
    T_F                              & tf        & \frac{1}{2}                        & \frac{1}{2}   & 1 \\[5pt]
    C_A                              & ca        & N                                  & 3             & 0 \\[5pt]
    C_F                              & cf        & \frac{N^2-1}{2N}                   & \frac{4}{3}   & 1 \\[8pt]
    \frac{d_A^{abcd}d_A^{abcd}}{N_A} & [d4AA/na] & \frac{N^2(N^2+36)}{24}             & \frac{135}{8} & 0 \\[8pt]
    \frac{d_F^{abcd}d_A^{abcd}}{N_A} & [d4FA/na] & \frac{N(N^2+6)}{48}                & \frac{15}{16} & 0 \\[8pt]
    \frac{d_F^{abcd}d_F^{abcd}}{N_A} & [d4FF/na] & \frac{N^4-6N^2+18}{96N^2}          & \frac{5}{96}  & 1 \\[8pt]
    \frac{d_F^{abcd}d_A^{abcd}}{N_R} & [d4FA/nr] & \frac{(N^2+6)(N^2-1)}{48}          & \frac{5}{2}   & 0 \\[8pt]
    \frac{d_F^{abcd}d_F^{abcd}}{N_R} & [d4FF/nr] & \frac{(N^4-6N^2+18)(N^2-1)}{96N^3} & \frac{5}{36}  & 1 \\[8pt]
    \hline
  \end{tabular}
  \caption{Symbols for colour factors.}
  \label{tab:colour}
\end{table}

\begin{table}[ht]
  \centering
  \renewcommand{\arraystretch}{1.2}
  \begin{tabular}{MTc|MT}
    \hline \rule{0pt}{3ex}
    \Pi^B              & gluonB & & a^B   & a \\
    \tilde{\Pi}^B      & ghostB & & \xi^B & xi \\
    \Sigma_V^B         & quarkB & \\
    T_i^B              & glglgl`i'B  & \verb|i| = \verb|1| or \verb|2| \\
    \tilde{\Gamma}_i^B & ghghgl`i'B  & \verb|i| = \verb|h| or \verb|g| \\
    \Lambda_i^B        & ququgl`i'B  & \verb|i| = \verb|q| or \verb|g| \\
    \Lambda_i^{T,B}    & ququglT`i'B & \verb|i| = \verb|q| or \verb|g| \\
    \hline
  \end{tabular}
  \caption{Symbols for unrenormalized results.}
  \label{tab:unrenorm}
\end{table}

\begin{table}[ht]
  \centering
  \renewcommand{\arraystretch}{1.2}
  \begin{tabular}{MTc|MT}
    \hline \rule{0pt}{3ex}
    \Pi^\MSbar              & gluonMSb & & a^\MSbar   & a \\
    \tilde{\Pi}^\MSbar      & ghostMSb & & \xi^\MSbar & xi \\
    \Sigma_V^\MSbar         & quarkMSb & \\
    T_i^\MSbar              & glglgl`i'MSb  & \verb|i| = \verb|1| or \verb|2| \\
    \tilde{\Gamma}_i^\MSbar & ghghgl`i'MSb  & \verb|i| = \verb|h| or \verb|g| \\
    \Lambda_i^\MSbar        & ququgl`i'MSb  & \verb|i| = \verb|q| or \verb|g| \\
    \Lambda_i^{T,\MSbar}    & ququglT`i'MSb & \verb|i| = \verb|q| or \verb|g| \\
    \hline
  \end{tabular}
  \caption{Symbols for renormalized results.}
  \label{tab:renorm}
\end{table}

\clearpage
\bibliographystyle{JHEP}
\bibliography{refs}

\end{document}